\documentstyle[prd,aps,preprint,tighten,epsfig]{revtex}

\begin{document}

\draft

\title{Direct CP Violation
and Isospin Triangles of $B \rightarrow \pi\pi$ Decays}
\author{\bf Zhi-zhong Xing}
\address{CCAST (World Laboratory), P.O. Box 8730, 
Beijing 100080, China \\ 
and Institute of High Energy Physics, 
Chinese Academy of Sciences, \\
P.O. Box 918 (4), Beijing 100039, China 
\footnote{Mailing address} \\
({\it Electronic address: xingzz@mail.ihep.ac.cn}) }
\maketitle

\begin{abstract}
The recent observation of $B^0_d \rightarrow \pi^0\pi^0$ and
$\bar{B}^0_d \rightarrow \pi^0\pi^0$ decay modes allows us to
make a fresh isospin analysis of $B\rightarrow \pi\pi$ 
transitions. We find that current experimental data can impose
some model-independent constraints on the parameter space of 
direct CP violation in $B_d \rightarrow \pi^+\pi^-$ and 
$B_d\rightarrow \pi^0\pi^0$. Furthermore, we establish 
a direct relationship between the weak phase $\alpha$ and
the charge-averaged branching fractions and CP-violating
asymmetries of $B\rightarrow \pi\pi$ decays.
\end{abstract}


\newpage

\vspace{1cm}

The major goal of $B$-meson factories is to test the 
Kobayashi-Maskawa mechanism of CP violation within the standard model 
and to detect possible new sources of CP violation beyond the
standard model. So far the CP-violating asymmetry in
$B^0_d$ vs $\bar{B}^0_d \rightarrow J/\psi K_{\rm S}$ decays has 
unambiguously been measured at KEK and SLAC \cite{2B}, and the 
experimental result is compatible very well with the standard-model 
expectation. Some preliminary evidence for CP violation in a number of
other $B$ decays has also been reported by BaBar and Belle 
Collaborations \cite{Yamamoto}. Further experiments will provide 
much more precise data on CP violation in the $B$-meson system, from 
which one may cross-check the consistency of the Kobayashi-Maskawa 
picture and probe possible new physics.

The charmless two-body nonleptonic decays 
$\bar{B}^0_d \rightarrow \pi^+ \pi^-$, 
$\bar{B}^0_d \rightarrow \pi^0 \pi^0$ and 
$B^-_u \rightarrow \pi^0 \pi^-$,
which can be related to one another via the isospin 
triangle \cite{Gronau}, have been of great interest in $B$ 
physics for a stringent test of the factorization hypothesis, a 
quantitative analysis of final-state interactions, and a clean 
determination of the CP-violating phase 
$\alpha \equiv \arg [-(V_{td}V^*_{tb})/(V_{ud}V^*_{ub})]$ 
(an inner angle of the well-known unitarity triangle of the 
Cabibbo-Kobayashi-Maskawa (CKM) quark mixing matrix \cite{PDG}). Among 
these three decay modes, $\bar{B}^0_d \rightarrow \pi^0 \pi^0$ is 
color-suppressed and should have the smallest branching fraction. The 
observation of this rare decay has recently been reported by 
BaBar \cite{BB} and Belle \cite{BE} Collaborations 
\footnote{Note that the experimentally-reported branching fractions of 
$B \rightarrow \pi\pi$ decays are all charge-averaged.}:
\begin{eqnarray}
{\cal B}_{00} & \equiv & \frac{{\cal B}(B^0_d \rightarrow \pi^0\pi^0)
+ {\cal B}(\bar{B}^0_d \rightarrow \pi^0\pi^0)}{2}
\nonumber \\
& = & \left \{ \matrix{
(2.1 \pm 0.6 \pm 0.3) \times 10^{-6} ~~ ({\rm BaBar}) \;  \cr
(1.7 \pm 0.6 \pm 0.3) \times 10^{-6} ~~ ({\rm Belle}) \; , ~ \cr} \right .
\end{eqnarray}
where the first error is statistical and the second one is systematic.
The experimental value of ${\cal B}_{00}$ is a bit larger than the
theoretical prediction in various QCD models \cite{Beneke}. In
comparison, the observed branching fractions of 
$\bar{B}^0_d \rightarrow \pi^+ \pi^-$ and 
$B^-_u \rightarrow \pi^0 \pi^-$ decays \cite{Fry}
\begin{eqnarray}
{\cal B}_{+-} & \equiv & \frac{{\cal B}(B^0_d \rightarrow \pi^+\pi^-)
+ {\cal B}(\bar{B}^0_d \rightarrow \pi^+\pi^-)}{2}
\nonumber \\
& = & \left \{ \matrix{
(4.7 \pm 0.6 \pm 0.2) \times 10^{-6} ~~ ({\rm BaBar}) \; \cr
(4.4 \pm 0.6 \pm 0.3) \times 10^{-6} ~~ ({\rm Belle}) \; ~ \cr
(4.5 \pm 1.4 \pm 0.5) \times 10^{-6} ~~ ({\rm CLEO}) \;  \cr} \right .
\end{eqnarray}
and
\begin{eqnarray}
{\cal B}_{0\pm ~} & \equiv & \frac{{\cal B}(B^+_u \rightarrow \pi^0\pi^+)
+ {\cal B}(B^-_u \rightarrow \pi^0\pi^-)}{2}
\nonumber \\
& = & \left \{ \matrix{
(5.5 \pm 1.0 \pm 0.6) \times 10^{-6} ~~ ({\rm BaBar}) \; \cr
(5.3 \pm 1.3 \pm 0.5) \times 10^{-6} ~~ ({\rm Belle}) \; ~ \cr
(4.6 \pm 1.8 \pm 0.7) \times 10^{-6} ~~ ({\rm CLEO}) \; \cr} \right .
\end{eqnarray}
are essentially consistent with the theoretical results \cite{Beneke}. 
Because ${\cal B}_{+-}$, ${\cal B}_{00}$ and ${\cal B}_{0\pm}$ are 
all charge-averaged, they cannot directly be related to one
another through the isospin relation. 

The purpose of this note is to make a fresh isospin analysis of three
$B\rightarrow \pi\pi$ decays by using new experimental data and
taking account of CP violation. We find that current data can impose
some useful and model-independent constraints on the parameter space 
of direct CP violation in $B_d \rightarrow \pi^+\pi^-$ and 
$B_d\rightarrow \pi^0\pi^0$ transitions. It is also possible to 
establish a direct relationship between the weak phase $\alpha$ and
the charge-averaged branching fractions and CP-violating
asymmetries of $B\rightarrow \pi\pi$ decays.

Let us define the {\it direct} CP-violating asymmetries between 
$\bar{B}^0_d \rightarrow \pi^+ \pi^-$, 
$\bar{B}^0_d \rightarrow \pi^0 \pi^0$,
$B^-_u \rightarrow \pi^0 \pi^-$ and their CP-conjugate decays:
\begin{eqnarray}
{\cal A}_{+-} & \equiv & \frac{{\cal B}(\bar{B}^0_d \rightarrow \pi^+\pi^-)
- {\cal B}(B^0_d \rightarrow \pi^+\pi^-)}
{{\cal B}(\bar{B}^0_d \rightarrow \pi^+\pi^-)
+ {\cal B}(B^0_d \rightarrow \pi^+\pi^-)} \;\; ,
\nonumber \\
{\cal A}_{00 ~} & \equiv & \frac{{\cal B}(\bar{B}^0_d \rightarrow \pi^0\pi^0)
- {\cal B}(B^0_d \rightarrow \pi^0\pi^0)}
{{\cal B}(\bar{B}^0_d \rightarrow \pi^0\pi^0)
+ {\cal B}(B^0_d \rightarrow \pi^0\pi^0)} \;\; ,
\nonumber \\
{\cal A}_{0\pm ~} & \equiv & \frac{{\cal B}(B^-_u \rightarrow \pi^0\pi^-)
- {\cal B}(B^+_u \rightarrow \pi^0\pi^+)}
{{\cal B}(B^-_u \rightarrow \pi^0\pi^-)
+ {\cal B}(B^+_u \rightarrow \pi^0\pi^+)} \;\; .
\end{eqnarray}
Taking account of Eqs. (1)--(4), one can easily obtain the magnitude 
of each decay amplitude: 
\begin{eqnarray}
\left |A(\bar{B}^0_d \rightarrow \pi^+\pi^-) \right | & \propto & 
\sqrt{\frac{{\cal B}_{+-}}{\tau^{~}_0} 
\left (1 + {\cal A}_{+-} \right )} \;\; ,
\nonumber \\
\left |A(\bar{B}^0_d \rightarrow \pi^0\pi^0) ~ \right | & \propto & 
\sqrt{\frac{{\cal B}_{00}}{\tau^{~}_0} 
\left (1 + {\cal A}_{00} \right )} \;\; ,
\nonumber \\
\left |A(B^-_u \rightarrow \pi^0\pi^-) \right | & \propto & 
\sqrt{\frac{{\cal B}_{0\pm}}{\tau^{~}_\pm} 
\left (1 + {\cal A}_{0\pm} \right )} \;\; ;
\end{eqnarray}
and 
\begin{eqnarray}
\left |A(B^0_d \rightarrow \pi^+\pi^-) \right | & \propto & 
\sqrt{\frac{{\cal B}_{+-}}{\tau^{~}_0} 
\left (1 - {\cal A}_{+-} \right )} \;\; ,
\nonumber \\
\left |A(B^0_d \rightarrow \pi^0\pi^0) ~ \right | & \propto & 
\sqrt{\frac{{\cal B}_{00}}{\tau^{~}_0} 
\left (1 - {\cal A}_{00} \right )} \;\; ,
\nonumber \\
\left |A(B^+_u \rightarrow \pi^0\pi^+) \right | & \propto & 
\sqrt{\frac{{\cal B}_{0\pm}}{\tau^{~}_\pm} 
\left (1 - {\cal A}_{0\pm} \right )} \;\; ,
\end{eqnarray}
where $\tau^{~}_0$ and $\tau^{~}_{\pm}$ denote the lifetimes
of neutral and charged $B$ mesons, respectively. The present
experimental data yield 
$\kappa \equiv \tau^{~}_0/\tau^{~}_{\pm} \approx 0.92$ \cite{PDG}. 
In Eqs. (5) and (6), we have neglected the tiny phase space 
difference between $\pi^0\pi^\pm$ and $\pi^+\pi^-$ (or 
$\pi^0\pi^0$) states. It is obvious that the relevant decay 
amplitudes cannot be determined, unless the CP-violating
asymmetries ${\cal A}_{+-}$, ${\cal A}_{00}$ and 
${\cal A}_{0\pm}$ are all measured.

Under isospin symmetry and in the neglect of electroweak 
penguin contributions \cite{Rosner}, the amplitudes of 
$\bar{B}^0_d \rightarrow \pi^+ \pi^-$, 
$\bar{B}^0_d \rightarrow \pi^0 \pi^0$ and
$B^-_u \rightarrow \pi^0 \pi^-$ decays form an isospin triangle 
in the complex plane:
\begin{equation}
A(\bar{B}^0_d \rightarrow \pi^+\pi^-) + \sqrt{2}
A(\bar{B}^0_d \rightarrow \pi^0\pi^0) \; = \;
\sqrt{2} A(B^-_u \rightarrow \pi^0 \pi^-) \; .
\end{equation}
Similarly,
\begin{equation}
A(B^0_d \rightarrow \pi^+\pi^-) + \sqrt{2}
A(B^0_d \rightarrow \pi^0\pi^0) \; = \;
\sqrt{2} A(B^+_u \rightarrow \pi^0 \pi^+) \; .
\end{equation}
In this approximation, the magnitudes of 
$A(B^-_u \rightarrow \pi^0 \pi^-)$ and
$A(B^+_u \rightarrow \pi^0 \pi^+)$ are identical to each
other \cite{Gronau}. Thus the CP-violating asymmetry
${\cal A}_{0\pm}$ vanishes. What we are concerned about is the
relative phase between the amplitudes of 
$\bar{B}^0_d \rightarrow \pi^+ \pi^-$
(or $B^0_d \rightarrow \pi^+ \pi^-$) and
$\bar{B}^0_d \rightarrow \pi^0 \pi^0$ 
(or $B^0_d \rightarrow \pi^0 \pi^0$) decays:
\begin{eqnarray}
\phi & \equiv & \arg \left [ \frac{A(\bar{B}^0_d \rightarrow \pi^+\pi^-)}
{A(\bar{B}^0_d \rightarrow \pi^0\pi^0)} \right ] \; ,
\nonumber \\
\varphi & \equiv & \arg \left [ \frac{A(B^0_d \rightarrow \pi^+\pi^-)}
{A(B^0_d \rightarrow \pi^0\pi^0)} \right ] \; .
\end{eqnarray}
At the tree level, $\varphi = \phi$ holds. A difference between
$\phi$ and $\varphi$ measures the QCD penguin effects in 
$B_d \rightarrow \pi^+ \pi^-$ and $B_d \rightarrow \pi^0 \pi^0$ 
transitions. With the help of Eqs. (5)--(8), we find
\begin{eqnarray}
\cos\phi & = & \frac{2 \kappa {\cal B}_{0\pm} - {\cal B}_{+-}
\left (1 + {\cal A}_{+-} \right ) - 
2 {\cal B}_{00} \left (1 + {\cal A}_{00} \right )}
{2 \sqrt{2 {\cal B}_{+-} {\cal B}_{00} \left (1 + {\cal A}_{+-} \right )
\left (1 + {\cal A}_{00} \right )}} \; ,
\nonumber \\
\cos\varphi & = & \frac{2 \kappa {\cal B}_{0\pm} - {\cal B}_{+-}
\left (1 - {\cal A}_{+-} \right ) - 
2 {\cal B}_{00} \left (1 - {\cal A}_{00} \right )}
{2 \sqrt{2 {\cal B}_{+-} {\cal B}_{00} \left (1 - {\cal A}_{+-} \right )
\left (1 - {\cal A}_{00} \right )}} \; ,
\end{eqnarray}
where ${\cal A}_{0\pm}=0$ has been taken into account. One can see
that $\varphi = \phi$ would hold, if both ${\cal A}_{+-}$ and
${\cal A}_{00}$ were vanishing. Hence the difference between
$\phi$ and $\varphi$ results from the penguin-induced CP violation 
in $B_d \rightarrow \pi^+ \pi^-$ and $B_d \rightarrow \pi^0 \pi^0$ 
decays.

To get a ball-park feeling of the discrepancy between $\phi$ and 
$\varphi$, we make use of the QCD factorization to estimate the 
ratio of $A(\bar{B}^0_d \rightarrow \pi^+ \pi^-)$ to
$A(\bar{B}^0_d \rightarrow \pi^0 \pi^0)$ and that of
$A(B^0_d \rightarrow \pi^+ \pi^-)$ to
$A(B^0_d \rightarrow \pi^0 \pi^0)$ \cite{Xing00}. The results are
\begin{eqnarray}
\frac{A(\bar{B}^0_d \rightarrow \pi^+\pi^-)}
{A(\bar{B}^0_d \rightarrow \pi^0\pi^0)}
& = & \frac{\left (a_1 + a^u_4 + a^u_6 \xi_\pi \right ) R_b e^{-i\gamma}
+ \left (a^c_4 + a^c_6 \xi_\pi \right )}
{\left (a_2 - a^u_4 - a^u_6 \xi_\pi \right ) R_b e^{-i\gamma}
- \left (a^c_4 + a^c_6 \xi_\pi \right )} \; ,
\nonumber \\
\frac{A(B^0_d \rightarrow \pi^+\pi^-)}
{A(B^0_d \rightarrow \pi^0\pi^0)}
& = & \frac{\left (a_1 + a^u_4 + a^u_6 \xi_\pi \right ) R_b e^{+i\gamma}
+ \left (a^c_4 + a^c_6 \xi_\pi \right )}
{\left (a_2 - a^u_4 - a^u_6 \xi_\pi \right ) R_b e^{+i\gamma}
- \left (a^c_4 + a^c_6 \xi_\pi \right )} \; ,
\end{eqnarray}
where $R_b \equiv |V_{ud}V_{ub}^*|/|V_{cd}V^*_{cb}|$ and
$\gamma \equiv \arg [-(V_{ud}V_{ub}^*)/(V_{cd}V^*_{cb})]$ are two 
parameters of the CKM unitarity triangle \cite{Buras03}; 
$a_1$, $a_2$, $a^{u,c}_4$ and $a^{u,c}_6$ denote the effective Wilson 
coefficients \cite{QCD}; $\xi_\pi$ is associated with
the penguin operators $Q_{5,6}$ \cite{Buras} and can be given as 
\begin{equation}
\xi_\pi \; =\; \frac{2 m^2_\pi}{(m_b - m_u) (m_u + m_d)}
\end{equation}
under isospin symmetry. For illustration, we typically adopt 
$R_b \approx 0.37$ and $\gamma \approx 65^\circ$ \cite{Buras03}
as well as $a_1 \approx 1.038+0.018i$, $a_2 \approx 0.082-0.080i$,
$a^u_4 \approx -0.029-0.015i$ and $a^c_4 \approx -0.034-0.008i$
at the scale $\mu = m_b$ \cite{QCD}. The formally power-suppressed
QCD coefficients $a^{u,c}_6$ can be neglected in the heavy quark
limit \cite{QCD}, because of $\xi_\pi \sim {\cal O}(1)$. Then we
arrive at $A(\bar{B}^0_d \rightarrow \pi^+\pi^-)/
A(\bar{B}^0_d \rightarrow \pi^0\pi^0) \approx 7.4 e^{-i17^\circ}$
and $A(B^0_d \rightarrow \pi^+\pi^-)/
A(B^0_d \rightarrow \pi^0\pi^0) \approx 4.3 e^{+i44^\circ}$. As
a result, $\phi \approx -17^\circ$ and $\varphi \approx 44^\circ$
are obtained. The big difference $\varphi - \phi \approx 61^\circ$
implies the existence of significant QCD penguin effects and large 
direct CP violation in $B_d \rightarrow \pi\pi$ decays. It should
be noted that $\phi$ and $\varphi$ are neither pure weak phases
nor pure strong phases, since the treel and penguin amplitudes of
each $B_d \rightarrow \pi\pi$ decay mode involve different
weak and strong phases. It should also be noted that
the numerical results for $\phi$ and $\varphi$ are
just illustrative, because possible final-state rescattering 
effects have not been taken into account in Eq. (11). 

Once the CP-violating asymmetries ${\cal A}_{+-}$ and ${\cal A}_{00}$
are measured at $B$ factories, it will be possible to determine
$\phi$ and $\varphi$ from Eq. (10) independently of the specific 
QCD models. A hint of ${\cal A}_{+-} \neq 0$ (of $\sim 2\sigma$ 
significance) has recently been reported by the Belle 
Collaboration, but the BaBar Collaboration's preliminary 
result for ${\cal A}_{+-}$ is consistent with zero \cite{Yamamoto}. 
There is no experimental information about ${\cal A}_{00}$ at
present. In this situation, we show the numerical correlation between
$\phi$ and $\varphi$ in Fig. 1(a) and that between ${\cal A}_{+-}$ 
and ${\cal A}_{00}$ in Fig. 1(b), where current data listed
in Eqs. (1)--(3) have been used. The points in Fig. 1(a) are 
generated from scanning the error bars of 
${\cal B}_{+-}$, ${\cal B}_{00}$ and ${\cal B}_{0\pm}$.
The light region in Fig. 1(b) is generated in a similar way
and corresponds to the allowed ranges of $\cos\phi$ and $\cos\varphi$
in Fig. 1(a), while the dark region in Fig. 1(b) results from 
inputting the central values
of ${\cal B}_{+-}$, ${\cal B}_{00}$ and ${\cal B}_{0\pm}$. 
Some comments and discussions are in order.

(1) Current experimental results of ${\cal B}_{+-}$, ${\cal B}_{00}$
and ${\cal B}_{0\pm}$ can impose some limited but model-independent
constraints on the values of $(\phi, \varphi)$ and 
$({\cal A}_{+-}, {\cal A}_{00})$. For example, the possibilities 
$\cos\phi = \cos\varphi = 0$ and 
$|{\cal A}_{+-}| = |{\cal A}_{00}| = 1$ are not allowed. 
As a consequence of $|{\cal A}_{+-}| <1$ and 
$|{\cal A}_{00}| <1$, the regions for $\cos\phi <-0.2$ and
$\cos\varphi < -02$ have completely been excluded by the present
experimental data.

(2) The error bars of ${\cal B}_{+-}$, ${\cal B}_{00}$ and 
${\cal B}_{0\pm}$ influence the parameter space of 
$(\phi, \varphi)$ or $({\cal A}_{+-}, {\cal A}_{00})$ to a
very limited extent. If ${\cal A}_{+-}$ is measured, one can
get a roughly allowed range of ${\cal A}_{00}$ from Fig. 1(b). 
For instance, ${\cal A}_{+-} \geq 0.6$ and
${\cal A}_{00} \geq 0.6$ cannot simultaneously hold. Nor can
${\cal A}_{+-} \leq -0.6$ and ${\cal A}_{00} \leq -0.6$
simultaneously hold.

(3) As the tree-level amplitude of $B^0_d\rightarrow \pi^0\pi^0$
or $\bar{B}^0_d\rightarrow \pi^0\pi^0$ is color-suppressed, its
magnitude might be comparable with the size of the corresponding
penguin amplitude. Then $|{\cal A}_{00}|$ is expected to be 
larger than $|{\cal A}_{+-}|$ in the QCD factorization. However,
the elastic rescattering effects in the final states of 
$B\rightarrow \pi\pi$ transitions could easily spoil this naive 
expectation. It is therefore useful to analyze the relevant
experimental data in a model-independent way.

(4) The validity of our numerical results relies on the validity 
of the isospin relations in Eqs. (7) and (8), which hold in the
assumption of negligible electroweak penguin effects.
The electroweak penguin contributions to 
$B\rightarrow \pi\pi$ transitions are generally expected to be 
insignificant \cite{Rosner}. This expectation would be problematic
or incorrect, if ${\cal A}_{0\pm} \neq 0$ were experimentally 
established. It is also worth mentioning that final-state
interactions in $B\rightarrow \pi\pi$ decays include both elastic 
$\pi\pi \rightleftharpoons \pi\pi$ rescattering and some 
possible inelastic rescattering effects 
\footnote{The amplitudes of $\bar{B}^0_d \rightarrow \pi^+ \pi^-$, 
$\bar{B}^0_d \rightarrow \pi^0 \pi^0$ and
$B^-_u \rightarrow \pi^0 \pi^-$ decay modes (or their CP-conjugate
processes) may still form an isospin 
triangle in the complex plane, even if the inelastic 
$\pi\pi \rightleftharpoons D\bar{D}$ rescattering effects are 
taken into account \cite{X95}. In this specific case, the 
analytical result obtained in Eq. (10) remains valid.
It is even expected that inelastic final-state interactions
could account for the existing discrepancy between experimental
data and QCD models for $D\rightarrow \pi\pi$ and 
$B\rightarrow \pi\pi$ decays \cite{Cheng}.}.
Whether inelastic final-state interactions are negligibly 
small or not remains an open question \cite{Donoghue}. To answer 
this question requires more precise data on both the branching 
fractions and the CP-violating asymmetries of 
$B\rightarrow \pi\pi$ decays.

We proceed to discuss the extraction of the CP-violating phase
$\alpha$ with the help of the isospin triangles in 
Eqs. (7) and (8). As shown in Ref. \cite{Gronau}, $\alpha$ 
depends on two CP-violating observables which arise from the 
interplay of decay and $B^0_d$-$\bar{B}^0_d$ mixing in 
$B_d\rightarrow \pi^+\pi^-$ and $B_d\rightarrow \pi^0\pi^0$
decays:
\begin{eqnarray}
\chi_{+-} \equiv \; {\rm Im} \left [ \frac{q}{p} \cdot
\frac{A(\bar{B}^0_d \rightarrow \pi^+\pi^-)}
{A(B^0_d \rightarrow \pi^+\pi^-)} \right ]
& = & |R| \sin [ 2 (\alpha + \Theta)] \; ,
\nonumber \\
\chi_{00} \; \equiv \; {\rm Im} \left [ \frac{q}{p} \cdot
\frac{A(\bar{B}^0_d \rightarrow \pi^0\pi^0)}
{A(B^0_d \rightarrow \pi^0\pi^0)} \right ] ~
& = & |\bar{R}| \sin [ 2 (\alpha + \bar{\Theta})] \; ,
\end{eqnarray}
where
\begin{eqnarray}
R & = & |R| e^{2i\Theta} \; =\; \frac{1 - \bar{r}}{1-r} \; ,
\nonumber \\
\bar{R} & = & |\bar{R}| e^{2i\bar{\Theta}} \; =\;
\frac{2+ \bar{r}}{2+r} \; ,
\end{eqnarray}
with $r = |r| e^{i\theta} \equiv A_0/A_2$ and 
$\bar{r} = |\bar{r}| e^{i\bar{\theta}} \equiv \bar{A}_0/\bar{A}_2$
being the ratios of $I=0$ and $I=2$ isospin amplitudes in
$B^0_d \rightarrow \pi^+\pi^-$ (or $\pi^0\pi^0$) and 
$\bar{B}^0_d \rightarrow \pi^+\pi^-$ (or $\pi^0\pi^0$) decays.
Following Ref. \cite{Du}, we obtain
\begin{eqnarray}
|r| & = & \sqrt{3(a+b) -2} \;\; ,
\nonumber \\
\theta & = & \pm \arccos \left [\frac{6b-3a-2}
{4\sqrt{3(a+b)-2}} \right ] \; ;
\end{eqnarray}
and
\begin{eqnarray}
|\bar{r}| & = & \sqrt{3(\bar{a}+\bar{b}) -2} \;\; ,
\nonumber \\
\bar{\theta} & = & \pm \arccos \left [\frac{6\bar{b}-3\bar{a}-2}
{4\sqrt{3(\bar{a}+\bar{b})-2}} \right ] \; ,
\end{eqnarray}
where
\begin{eqnarray}
a & \equiv & \left | \frac{A(B^0_d \rightarrow \pi^+\pi^-)}
{A(B^+_u\rightarrow \pi^0\pi^+)} \right |^2 \; =\;
\frac{1}{\kappa} \cdot \frac{{\cal B}_{+-}}{{\cal B}_{0\pm}}
\left (1 - {\cal A}_{+-} \right ) \; ,
\nonumber \\
b & \equiv & \left | \frac{A(B^0_d \rightarrow \pi^0\pi^0)}
{A(B^+_u\rightarrow \pi^0\pi^+)} \right |^2 \; =\;
\frac{1}{\kappa} \cdot \frac{{\cal B}_{00}}{{\cal B}_{0\pm}}
\left (1 - {\cal A}_{00} \right ) \; ;
\end{eqnarray}
and
\begin{eqnarray}
\bar{a} & \equiv & \left | \frac{A(\bar{B}^0_d \rightarrow \pi^+\pi^-)}
{A(B^-_u\rightarrow \pi^0\pi^-)} \right |^2 \; =\;
\frac{1}{\kappa} \cdot \frac{{\cal B}_{+-}}{{\cal B}_{0\pm}}
\left (1 + {\cal A}_{+-} \right ) \; ,
\nonumber \\
\bar{b} & \equiv & \left | \frac{A(\bar{B}^0_d \rightarrow \pi^0\pi^0)}
{A(B^-_u\rightarrow \pi^0\pi^-)} \right |^2 \; =\;
\frac{1}{\kappa} \cdot \frac{{\cal B}_{00}}{{\cal B}_{0\pm}}
\left (1 + {\cal A}_{00} \right ) \; .
\end{eqnarray}
Clearly $r$ (or $\bar{r}$) can be determined up to a twofold
ambiguity in the sign of its phase, if both ${\cal A}_{+-}$ and
${\cal A}_{00}$ are measured. One can in turn determine $R$ 
(or $\bar{R}$), both its magnitude and its phase. It is then
possible to extract the weak phase $\alpha$ from Eq. (13)
through the time-dependent measurement of $\chi_{+-}$ and (or)
$\chi_{00}$ in $B_d \rightarrow \pi^+\pi^-$ and (or) 
$B_d \rightarrow \pi^0\pi^0$ decays. On the $\Upsilon (4S)$
resonance, the time-dependent rate asymmetry between $B^0_d(t)$ 
and $\bar{B}^0_d(t)$ decays into a CP eigenstate $f$ reads as
\begin{eqnarray}
\Delta_f (t) & \equiv & 
\frac{\Gamma [\bar{B}^0_d(t) \rightarrow f] -
\Gamma [B^0_d(t) \rightarrow f]}
{\Gamma [\bar{B}^0_d(t) \rightarrow f] +
\Gamma [B^0_d(t) \rightarrow f]}
\nonumber \\
& = & \frac{|\rho^{~}_f|^2 - 1}{|\rho^{~}_f|^2 + 1}
\cos (\Delta M_d \cdot t) ~ + 
\frac{\displaystyle 2 {\rm Im} \left (\frac{q}{p} \rho^{~}_f \right )}
{|\rho^{~}_f|^2 + 1} \sin (\Delta M_d \cdot t) \; , 
\end{eqnarray}
where $\rho^{~}_f \equiv A(\bar{B}^0_d \rightarrow f)/
A(B^0_d\rightarrow f)$ and tiny CP violation in $B^0_d$-$\bar{B}^0_d$
mixing has been neglected (i.e., $|q/p| \approx 1$) \cite{Xing96}. 
For $f=\pi^+\pi^-$ and $\pi^0\pi^0$, we take account of 
Eqs. (4)--(6) and (13) and obtain 
\begin{eqnarray}
\Delta_{\pi^+\pi^-}(t) & = & {\cal A}_{+-} \cos (\Delta M_d \cdot t) 
 +  (1 - {\cal A}_{+-}) \chi_{+-} \sin (\Delta M_d \cdot t) \; , 
\nonumber \\
\Delta_{\pi^0\pi^0}(t) ~ & = & {\cal A}_{00} \cos (\Delta M_d \cdot t) 
~ + ~ (1 - {\cal A}_{00}) \chi_{00} \sin (\Delta M_d \cdot t) \; .
\end{eqnarray}
Clearly both ${\cal A}_{+-}$ (or ${\cal A}_{00}$) 
and $\chi_{+-}$ (or $\chi_{00}$) can be determined from measuring 
the time distribution of $\Delta_{\pi^+\pi^-}$ 
(or $\Delta_{\pi^0\pi^0}$) at asymmetric $B$ factories. Compared with 
Refs. \cite{Gronau,Du}, the formulas obtained in Eqs. (13)--(18) 
and (20) establish a more {\it direct} relationship between 
$\alpha$ and the observables ${\cal B}_{+-}$, ${\cal B}_{00}$,
${\cal B}_{0\pm}$, ${\cal A}_{+-}$ and ${\cal A}_{00}$. Thus
the usefulness of our results is noteworthy.

To summarize, we have presented a fresh isospin analysis of 
rare $B\rightarrow \pi\pi$ decays by taking account of the fact 
that the experimentally-reported branching fractions are
charge-averaged and large CP violation may exist in them.
We find that current experimental data can impose some useful
and model-independent constraints on the parameter space of
direct CP violation in $B_d\rightarrow \pi^+\pi^-$ and
$B_d\rightarrow \pi^0\pi^0$ transitions. A straightforward 
relationship between the CP-violating phase $\alpha$ and
the measurables of $B\rightarrow \pi\pi$ decays has also
been established. More precise measurements of such charmless
$B$ decays will allow us to test the consistency of the
Kobayashi-Maskawa mechanism of CP violation and probe possible
new physics beyond the standard model.

\vspace{0.4cm}

The author likes to thank Y.F. Zhou for useful discussions and
partial involvement at the very early stage of this work 
(in 2002). This research was supported in part by the National 
Nature Science Foundation of China.

\newpage

\begin{figure}[t]
\vspace{-1cm}
\epsfig{file=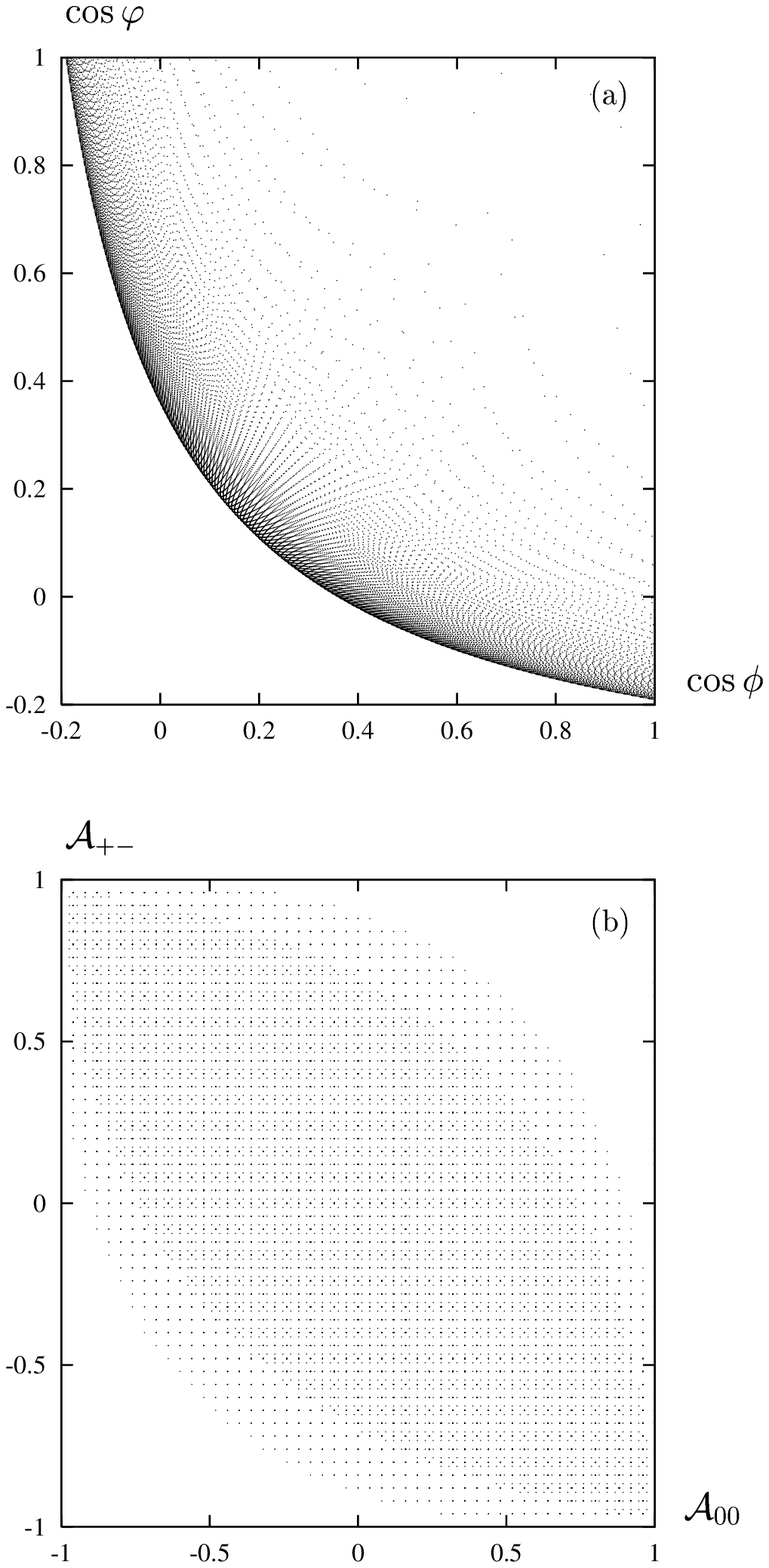,bbllx=2.3cm,bblly=4.3cm,bburx=18cm,bbury=30cm,%
width=14cm,height=20cm,angle=0,clip=0}
\vspace{-0cm}
\caption{Constraints of current experimental data on (a) the 
parameter space of $\cos\phi$ and $\cos\varphi$ and (b) the parameter
space of ${\cal A}_{00}$ and ${\cal A}_{+-}$. Note that the dark region 
of (b) corresponds to the central values of ${\cal B}_{+-}$,
${\cal B}_{00}$ and ${\cal B}_{0\pm}$ given in Eqs. (1)--(3).}  
\end{figure}

\end{document}